\documentclass[conference]{IEEEtran}

\usepackage{algorithm}
\usepackage{algorithmic}

\IEEEoverridecommandlockouts
\usepackage{cite}
\usepackage{amsmath,amssymb,amsfonts}
\usepackage{bbm}
\usepackage{graphicx}
\usepackage{textcomp}
\usepackage{xcolor}
\usepackage{acronym}

\allowdisplaybreaks
\def\BibTeX{{\rm B\kern-.05em{\sc i\kern-.025em b}\kern-.08em
    T\kern-.1667em\lower.7ex\hbox{E}\kern-.125emX}}
\graphicspath{{figures/}}

\acrodef{HARQ}{Hybrid Automatic ReQuest}
\acrodef{MIMO}{Multiple Input-Multiple Output}
\acrodef{TDMA}{Time Division Multiple Access}
\acrodef{OFDMA}{Orthogonal Frequency Division Multiple Access}
\acrodef{NOMA}{Non-Orthogonal Multiple Access}
\acrodef{CDMA}{Code Division Multiple Access}
\acrodef{SIC}{Successive Interference Canceler}
\acrodef{SNR}{Signal-to-Noise Ratio}
\acrodef{ACK}{Acknowlegment}
\acrodef{NACK}{Nonacknowledgement}

\begin{document}

\title{Analysis of Multi-Messages Retransmission Schemes}
\author{
  \IEEEauthorblockN{Alaa Khreis\textsuperscript{1}, Francesca Bassi\textsuperscript{2},~Philippe Ciblat\textsuperscript{3},~Pierre Duhamel\textsuperscript{4}}
\IEEEauthorblockA{
	\textsuperscript{1} Huawei, Boulogne-Billancourt, France\\
	\textsuperscript{2} IRT SystemX, Palaiseau, France\\
	\textsuperscript{3} Telecom ParisTech, Institut Polytechnique de Paris, Paris, France\\
	\textsuperscript{4} L2S, CNRS, Gif-sur-Yvette, France\\
        philippe.ciblat@telecom-paristech.fr}}
\maketitle
\begin{abstract}
  Hybrid Automatic ReQuest (HARQ) protocol enables reliable communications in wireless systems. Usually, several parallel streams are sent in successive timeslots following a time-sharing approach. Recently, multi-layer HARQ has been proposed by superposing packets within a timeslot. In this paper, we evaluate the potential of this multi-layer HARQ by playing with some design parameters. We show that a gain in throughput is only obtained at mid-Signal-to-Noise Ratio (SNR).    
\end{abstract}

\section{Introduction}
\let\thefootnote\relax\footnotetext{\noindent -----------------\\ This work was supported by Labex Digicosme under the grant ``Coccinelle''.}

In order to support the increase data rate demand, current wireless communication systems have to manage properly {\it i)} the channel fading and {\it ii)} the multi-messages resource sharing (if each message belongs to a specific user, this boils down to the multiple access technique). The first drawback is mitigated with diversity technique, such as retransmission (via \ac{HARQ}) or \ac{MIMO}. The second drawback is so far handled with orthogonal transmission techniques such as \ac{TDMA} (also called Time-sharing) or \ac{OFDMA}. Only recently, some non-orthogonal techniques such as \ac{NOMA} are envisioned for future systems. Nevertheless, multi-user interference may occur even in an orthogonal systems when asynchronism is encountered (see \ac{CDMA} in 3G) or partial frequency reuse for multi-cell environment is considered. Then multi-user aware receivers are carried out such as \ac{SIC}.

Parallel \ac{HARQ} is a way to implement \ac{HARQ} when the feedback is delayed. But once again, only one packet related to one message is sent in each timeslot. Nevertheless some works have proposed \cite{shamai08, ensea09,Szczecinski15,khreis} to use \ac{HARQ} with a non-orthogonal combination of different messages, typically, with a superposition coding technique. In other words, a sum of packets related to different messages is sent at each timeslot. Therefore this approach is also called multi-layer \ac{HARQ}. 
These works have shown that the multi-layer \ac{HARQ} has a great potential over different system settings.
For instance, in \cite{shamai08}, the multi-layer approach is mainly analyzed with a constant channel over the timeslots preventing from clearly exhibiting the possible diversity gain. Moreover when a new transmission is triggered, there is no superposition with previous packets. In \cite{ensea09}, the multi-layer approach proposed by \cite{shamai08} is evaluated with real coding schemes. In \cite{Szczecinski15}, the next timeslot is shared in time or by superposition according to the state of the communication (channel, accumulated mutual information). In \cite{khreis}, a new multi-layer \ac{HARQ} approach is developed by taking into account the feedback delay. All these results have been obtained by assuming a constant transmission rate $R$ while since \cite{jindal} we know that this parameter may strongly modify the behavior of \ac{HARQ} protocol.

In this paper we analyze a slice of a parallel \ac{HARQ} protocol when superposition coding is added at each timeslot. We also compare it with the raw superposition coding when \ac{HARQ} is dropped. We so analyze the trade-off between diversity (provided by \ac{HARQ}) and multiplexing (provided by superposition coding) by playing with different hyper-parameters such as the rate, the power proportion amongst the superposed packets.

The paper is organized as follows: in Section \ref{sec:model} we introduce the system model and the different analyzed \ac{HARQ} protocols. In Section \ref{sec:T}, we derive analytically the throughput for each analyzed protocol. Numerical results with an analysis of the pros and cons of each protocol are done in Section \ref{sec:simus}. Concluding remarks are drawn in Section \ref{sec:ccl}.

\section{System model}\label{sec:model}
We consider a parallel \ac{HARQ} protocol where the timeslots $t$ and $kT+t$ (with $k\in\mathbb{N}^\star$) are devoted the user/stream $u_0$. The receiver related to user $u_0$ attempts to decode the involved current message at the end of each timeslot $t_0$ (devoted to user $u_0$)  and a feedback taking the value \ac{ACK} or \ac{NACK} for the involved message is received without error before the beginning of timeslot $T+t_0$ at the transmitter side. This standard parallel \ac{HARQ} is hereafter called ``{\it Time-Sharing}'' (TS) approach since each timeslot is devoted to one packet associated with one message.

\begin{figure}[htb]
\begin{center}
\includegraphics[width=0.8\columnwidth]{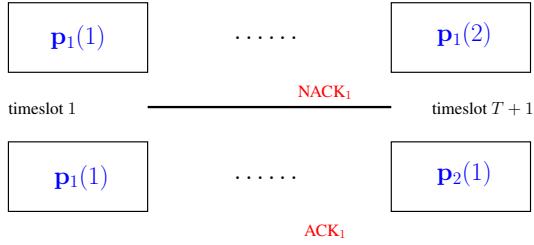}
\caption{Slice of standard parallel \ac{HARQ}}\label{fig:config1}
\end{center}
\end{figure}

In Fig.~\ref{fig:config1}, we draw a slice of two consecutive timeslots devoted to user $u_0$. The white space between both timeslots enable to insert the other \ac{HARQ} protocols devoted t oother uses or stream in the framework of parallel \ac{HARQ}.
More precisely, packet $\mathbf{p}_1(1)$ (related to message $m_1$) is sent at timeslot $1$ with full power $P$. If a NACK is received before timeslot $T+1$, packet $\mathbf{p}_1(2)$ (still related to message $m_1$) is sent at time slot $T+1$. If an ACK is received before timeslot $T+1$, packet $\mathbf{p}_2(1)$  (related to a new message $m_2$) is sent at timeslot $T+1$. Packets are sent with full power $P$.

\begin{figure}[htb]
\begin{center}
\includegraphics[width=0.8\columnwidth]{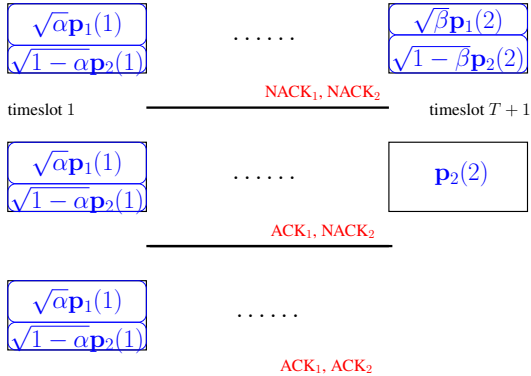}
\caption{Slice of multi-layer \ac{HARQ}}\label{fig:config2}
\end{center}
\end{figure}

Based on the standard parallel \ac{HARQ} described in Fig.~\ref{fig:config1}, seen as the first layer, we propose to add a second layer managing the packet not currently sent by the first layer. This approach is called ``{\it multi-layer \ac{HARQ}}''. Once again, we consider a slice of two consecutive timeslots devoted to the same user $_0$. In timeslot $1$, we send a linear combination of packets related to $m_1$ and $m_2$ at powers $\alpha P$ and $(1-\alpha)P$ respectively. After the reception of timeslot $1$, the receiver attempts to decode both messages. In timeslot $T+1$:
\begin{itemize}
\item if both messages are not decoded, the transmitter sends a linear combination of packets related to $m_1$ and $m_2$ with $\beta P$ and $(1-\beta)P$ respectively.
\item if one message is decoded, the receiver removes it and the transmitter sends the other one at full power $P$.
\item if both messages are decoded, the transmitter transmits new messages $m_3$ adn $m_4$.
\end{itemize}
The idea of multi-layer \ac{HARQ} is summarized in Fig.~\ref{fig:config2}.

\begin{figure}[htb]
\begin{center}
\includegraphics[width=0.8\columnwidth]{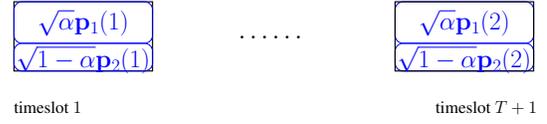}
\caption{Slice of superposition coding}\label{fig:config3}
\end{center}
\end{figure}

Instead of sending two packets into two timeslots by assigning one packet per timeslot, we can envisioned to apply the approach ``Superposition coding'' (SC) over two timeslots. In that case, in both timeslots (so there is no feedback at the end of the first timeslot), we send a linear combination of packets related to $m_1$ and $m_2$. We consider powers $\alpha P$ and $(1-\alpha)P$ for packets $1$ and $2$ respectively. The feedback for both messages is sent at the end of the second timeslot as summarized in Fig.~\ref{fig:config3}.


Note that the channel gain at timeslot $1$ (resp. $2$) is $g_1$ (resp. $g_2$). But the variance is identical. We denote $\sigma^2=\mathbb{E}[g_1]=\mathbb{E}[g_2]$. Noise is assumed to be unit-variance. The duration of each timeslot is $N$, and the number of information bits associated with each message is $RN$. Note also that we focus on the analysis of the elementary slice of two consecutive timeslots as drawing in the previous figures.


\section{Throughput derivations}\label{sec:T}
The objective of the section is to express the throughput for any previously-mentioned \ac{HARQ} protocols in closed-form.

In order to calculate these throughputs, we need to characterize all the below-listed events leading to at least one correct message decoding.

\begin{itemize}
\item Event $\Omega_0$: $(m_1,m_2)$ decoded at timeslot 1. $2RN$ received information bits,
\item Event $\Omega_1$: only $m_1$ decoded at timeslot 1, and $m_2$  decoded at timeslot $2$. $2RN$ received information bits, 
\item Event $\Omega_1'$: only $m_2$ decoded at timeslot 1, and $m_1$  decoded at timeslot $2$. $2RN$ received information bits,
\item Event $\Omega_2$: only  $m_1$ decoded at timeslot 1, and $m_2$ not decoded at timeslot $2$. $RN$ received information bits,
\item Event $\Omega_2'$:  only $m_2$ decoded at timeslot 1, and $m_1$ not decoded at timeslot $2$. $RN$ received information bits,
\item Event $\Omega_3$:  $(m_1,m_2)$ not decoded at timeslot 1, and $(m_1,m_2)$ decoded at timeslot $2$. $2RN$ received information bits,
\item Event $\Omega_4$:  $(m_1,m_2)$ not decoded at timeslot 1, and only $m_1$ decoded at timeslot $2$. $RN$ received information bits,
\item Event $\Omega_4'$:  $(m_1,m_2)$ not decoded at timeslot 1, and only $m_2$ decoded at timeslot $2$. $RN$ received information bits.
\end{itemize}

We denote the probability of the event $\Omega_0$ by $P_0(\alpha)$. For $i\in\{1, 2\}$, we denote the probability of event $\Omega_i$ by $\mathcal{P}_i(\alpha)$ and the probability of event $\Omega_i'$ by $\mathcal{P}_i'(\alpha)$. They only depend on $\alpha$ because  one packet is well acknowledged in the first timeslot which implies that the second timeslot is not shared and the total power is given to the remaining message. For $i\in\{3, 4\}$, we denote the probability of event $\Omega_i$ by $\mathcal{P}_i(\alpha,\beta)$ and the probability of event $\Omega_4'$ by $\mathcal{P}_4'(\alpha,\beta)$. 

Before going further, we write the throughput, denoted by $\eta$, for the different HARQ protocols with respect to $\mathcal{P}_i$ for $i\in\{0,1,2,3,4\}$ and  $\mathcal{P}_i'$ for $i\in\{1,2,4\}$.

\paragraph{Protocol ``Time-Sharing''}
by applying renewal theory, we know that the numerator is the average reward (correctly received bits) and the denominator is the average time used by each state. Therefore
\begin{eqnarray}
 \nonumber  \eta_{\textrm{TS}}&=& \frac{  2RN\mathcal{P}_1(1)  + RN\mathcal{P}_2(1) + RN \mathcal{P}_4(1,1)}{2N}\\
  &=& R\mathcal{P}_1(1)  + (R/2)(\mathcal{P}_2(1) +\mathcal{P}_4(1,1))
  \end{eqnarray}


\paragraph{Protocol ``Multi-layer HARQ''} by doing similar reasoning as the ``Time-Sharing'' protocol, we obtain
\begin{equation*}
  \eta_{\textrm{MLH}}= \frac{RNQ}{ N\mathcal{P}_0(\alpha) + 2N(1-\mathcal{P}_0(\alpha)) } 
\end{equation*}
with
\begin{eqnarray*}
  Q&=&2\mathcal{P}_0(\alpha)+ 2(\mathcal{P}_1(\alpha)+\mathcal{P}_1'(\alpha)) \\
   &+& 2\mathcal{P}_3(\alpha,\beta) +\mathcal{P}_2(\alpha)+\mathcal{P}_2'(\alpha)\\
  &+&\mathcal{P}_4(\alpha,\beta)+\mathcal{P}_4'(\alpha,\beta).
\end{eqnarray*}
Consequently, we have 
\begin{equation}
\eta_{\textrm{MLH}} = \frac{RQ}{ \mathcal{P}_0(\alpha) + 2(1-\mathcal{P}_0(\alpha)) }.
\end{equation}

Notice that $\eta_{\textrm{TS}}=\eta_{\textrm{MLH}}$ by forcing $\alpha=1$ and $\beta=1$ since $\mathcal{P}_0(1)= \mathcal{P}_1'(1)=\mathcal{P}_3(1,1)=\mathcal{P}_2'(1)= \mathcal{P}_4'(1,1)=0$.



\paragraph{Protocol ``Superposition coding''} for this protocol, at the end of the first timeslot, we do not attempt to decode the packets. Therefore the throughput does not write anymore with respect to the elementary probabilities described above. Actually, this throughput requires the following events $\tilde{\Omega}_3$, $\tilde{\Omega}_4$, $\tilde{\Omega}_4'$ where  the event $\tilde{\Omega}_i$ only corresponds to the last subevent describing $\Omega_i$. Then we define the probabilities of these events as follows: $\tilde{\mathcal{P}}_3(\alpha)$, $\tilde{\mathcal{P}}_3(\alpha)$, and $\tilde{\mathcal{P}}_4'(\alpha)$. Notice that these probabilities only depend on $\alpha$ whatever $\beta$. Then
\begin{eqnarray}
\nonumber  \eta_{\textrm{SC}} &=& \frac{2RN\tilde{\mathcal{P}}_3(\alpha)  + RN(\tilde{\mathcal{P}}_4(\alpha)+\tilde{\mathcal{P}}_4'(\alpha))}{2N}\\
& =& R\tilde{\mathcal{P}}_3(\alpha)  + (R/2)(\tilde{\mathcal{P}}_4(\alpha)+\tilde{\mathcal{P}}_4'(\alpha)).
\end{eqnarray}

\subsection{Closed-form expression for $\mathcal{P}_0$}
The event $\Omega_0$ corresponds to the case where $(m_1,m_2)$ are jointly  decoded at timeslot 1. It corresponds to the event of success for two messages within one timeslot when the following observation is available
$$ \mathbf{y}(1) =  \sqrt{\alpha P} g_1\mathbf{p}(1) +   \sqrt{(1-\alpha )P} g_1  \mathbf{p}(2)  + \textrm{noise}. $$
The MAC region of this channel (corresponding to success for packets 1 and 2) leads to the three equations \cite{mac}
$$\left\{\begin{array}{lcl} R&\leq&  \log(1+g_1\alpha P)  \\R&\leq& \log(1+g_1(1-\alpha)P)\\ 2R&\leq & \log(1+g_1P) \end{array}\right. . $$
We have to derive 
\begin{eqnarray*}
  \mathcal{P}_0(\alpha)&=&\textrm{Pr}( R \leq  \log(1+g_1\alpha P ),\\ && R\leq\log(1+g_1(1-\alpha)P),2R\leq\log(1+g_1P))\\
   &=&\textrm{Pr}(g_1 \geq \max\left(\frac{2^R-1}{\alpha P},\frac{2^R-1}{(1-\alpha) P},\frac{2^{2R}-1}{P}\right)).
\end{eqnarray*}
Finally, we have
\begin{equation}
\mathcal{P}_0(\alpha)= \frac{1}{\sigma^2}\int_{G_{\min}}^\infty   e^{-\frac{g}{\sigma^2}}dg  
\end{equation}
with
$$G_{\min}=  \max\left(\frac{2^R-1}{\alpha P},\frac{2^R-1}{(1-\alpha) P},\frac{2^{2R}-1}{P}\right).$$



\subsection{Closed-form expression for $\mathcal{P}_1$ and $\mathcal{P}_1'$ }
The event $\Omega_1$ corresponds to the case where in the first timeslot, the first packet is correctly received while the second one is not. In the second timeslot, the second packet (which is alone now due to the SIC applied on the first packet, but with power $P$) is correctly decoded.

In the first timeslot, the rate satisfies \cite{snc12}
\begin{equation}\label{t1-p1}
  \left\{\begin{array}{lcl} 
R&\leq&\log(1+\frac{g_1\alpha P}{1+g_1(1-\alpha)P})    \\
R&\geq&  \log(1+g_1(1-\alpha) P) \end{array}\right. .
  \end{equation}

Then in the second timeslot, we have $m_2$ which is alone and the rate is smaller than the accumulated mutual information over both timeslots. So 
$$R\leq  \log(1+g_1(1-\alpha) P) +  \log(1+g_2 P).  $$

Therefore
\begin{eqnarray*}
  \mathcal{P}_{1}(\alpha)&=&\textrm{Pr}(R\leq\log(1+\frac{g_1\alpha P}{1+g_1(1-\alpha)P}), \\ && R\geq  \log(1+g_1(1-\alpha) P),\\ & & R \leq \log(1+g_1(1-\alpha) P) +  \log(1+g_2P)  )\\
&=& \textrm{Pr}(2^R-1 \leq \frac{g_1\alpha P}{1+g_1(1-\alpha)P} , g_1\leq  \frac{2^R-1}{(1-\alpha) P},\\
& & 2^R \leq (1+g_1(1-\alpha) P)(1+g_2 P)  ).
\end{eqnarray*}

If $\alpha\leq (2^R-1)/(2^R)$, the first constraint is never true (not power enough to decode message $1$), and $\mathcal{P}_1(\alpha)=0$. Otherwise, we get 
\begin{eqnarray*}
  \mathcal{P}_{1}(\alpha)&=& \textrm{Pr}(g_1\geq \frac{2^R-1}{(1+2^R(\alpha-1) )P} , g_1\leq  \frac{2^R-1}{(1-\alpha) P},\\
& & 2^R \leq (1+g_1(1-\alpha) P)(1+g_2 P)  ).
\end{eqnarray*}
The set of feasible points $g_1$ is not empty if its lower-bound is smaller than its upper-bound. This is true when $\alpha\geq 2^R/(2^R+1)$. Moreover as  $2^R/(2^R+1)\geq  (2^R-1)/2^R$, we obtain that $\mathcal{P}_1(\alpha)=0$ when $\alpha\leq 2^R/(2^R+1)$. Otherwise, we have
\begin{eqnarray*}
  \mathcal{P}_1(\alpha)&=&\textrm{Pr}(\frac{2^R-1}{(1+2^R(\alpha-1) )P} \leq g_1\leq  \frac{2^R-1}{(1-\alpha) P},\\ &&g_2\geq \left(\frac{2^R}{1+g_1(1-\alpha) P}-1\right)^+ \frac{1}{ P})\end{eqnarray*}
where $(x)^+=\max(0,x)$. This operator has been added on the lower-bound of $g_2$ to ensure its positivity.
We deduce that
  \begin{eqnarray*}
    \mathcal{P}_1(\alpha)    &=& \frac{1}{\sigma^2} \int_{\frac{2^R-1}{(1+2^R(\alpha-1) )P} }^{\frac{2^R-1}{(1-\alpha) P}}  e^{- \frac{1}{\sigma^2  P}\left(\frac{2^R}{1+g(1-\alpha) P}-1 \right)^+}    e^{-\frac{g}{\sigma^2}} dg.
  \end{eqnarray*}


As for $\mathcal{P}_1'$, we just have to replace $\alpha$ with $(1-\alpha)$. So
$$\mathcal{P}_1'(\alpha)=\mathcal{P}_1(1-\alpha).$$


\subsection{Closed-form expressions for $\mathcal{P}_2$ and $\mathcal{P}_2'$}
The event $\Omega_2$ corresponds to the case where in the first timeslot, the first packet is correctly received while the second one is not. In the second timeslot, the second packet (which is alone now due to the SIC applied on the first packet, but with power $P$) is still not decoded.

The constraint on the rate for the first timeslot is still Eq.~\eqref{t1-p1}. 
Then in the second timeslot, we have $m_2$ which is alone and the rate is higher than the accumulated mutual information over both timeslots. So 
$$R\geq  \log(1+g_1(1-\alpha) P) +  \log(1+g_2 P).$$
Therefore
\begin{eqnarray*}
  \mathcal{P}_{2}(\alpha)\hspace{-2mm}&=&\hspace{-2mm}\textrm{Pr}(R\leq\log(1+\frac{g_1\alpha P}{1+g_1(1-\alpha)P}), \\ \hspace{-2mm}& & \hspace{-2mm}R\geq  \log(1+g_1(1-\alpha) P),\\ \hspace{-2mm}& &\hspace{-2mm} R \geq \log(1+g_1(1-\alpha) P) +  \log(1+g_2(1-\beta) P)  ).\end{eqnarray*}
It is easy to prove that
\begin{eqnarray*}
  \mathcal{P}_1(\alpha)+\mathcal{P}_2(\alpha)&=& \textrm{Pr}(R\leq\log(1+\frac{g_1\alpha P}{1+g_1(1-\alpha)P}), \\ &&R\geq  \log(1+g_1(1-\alpha) P)).
\end{eqnarray*}

Then, by applying the same reasoning as for $\mathcal{P}_1$, we obtain that $\mathcal{P}_2(\alpha)=0$ for $\alpha\leq 2^R/(2^R+1)$. Otherwise, we have
\begin{eqnarray*}
    \mathcal{P}_2(\alpha)\hspace{-2mm}&=&\hspace{-2mm}\textrm{Pr}(\frac{2^R-1}{(1+2^R(\alpha-1) )P} \leq g_1\leq  \frac{2^R-1}{(1-\alpha) P})-\mathcal{P}_1(\alpha)\\
    &=&   \left(  e^{- \frac{2^R-1}{(1+2^R(\alpha-1) )\sigma^2 P} } - e^{-\frac{2^R-1}{(1-\alpha)\sigma^2 P} }  \right)   - \mathcal{P}_1(\alpha).
\end{eqnarray*}

As for $\mathcal{P}_2'$, we just have to replace $\alpha$ with $(1-\alpha)$. So
$$\mathcal{P}_2'(\alpha)=\mathcal{P}_2(1-\alpha).$$


\subsection{Closed-form expression for $\mathcal{P}_3$}
The event $\Omega_3$ corresponds to the case where $(m_1,m_2)$ are not decoded at timeslot 1, and then $(m_1,m_2)$ are jointly decoded at timeslot $2$. Consequently 
$$\mathcal{P}_3=\textrm{Pr}(\omega_3,\omega_3')$$
where
\begin{itemize}
\item $\omega_3$ is the event of failure for two packets during the timeslot 1, i.e., when the observations are
$$ \mathbf{y}(1) =  \sqrt{\alpha P} g_1\mathbf{p}(1)  + \sqrt{(1-\alpha )P} g_1 \mathbf{p}(2)  + \textrm{noise}, $$
which leads to the following inequalities
 $$\left\{\begin{array}{lcl} 
            R&\geq& \log(1+\frac{g_1\alpha P }{1+g_1(1-\alpha) P  })  \\
R&\geq& \log(1+\frac{g_1(1-\alpha)P }{1+g_1\alpha P  })\\
2R&\geq & \log(1+g_1P)  \end{array}\right. $$

\item $\omega_3'$ is the event of success for two packets when the available observations are as follows 
$$ \left[\begin{array}{c} \mathbf{y}(1) \\ \mathbf{y}(2) \end{array} \right] =  \left[\begin{array}{cc} \sqrt{\alpha P} g_1 & \sqrt{(1-\alpha) P} g_1 \\ \sqrt{\beta P} g_2 & \sqrt{(1-\beta)P}g_2 \end{array} \right] \left[\begin{array}{c} \mathbf{p}(1) \\ \mathbf{p}(2) \end{array} \right]    + \textrm{noise}. $$

The MAC region of this channel (corresponding to success for packets 1 and 2) leads to the three equations 
$$\left\{\begin{array}{lcl} 
           R&\leq&  \log(1+g_1\alpha P) + \log(1+g_2 \beta P) \\
R&\leq&  \log(1+g_1(1-\alpha) P)   + \log(1+g_2(1-\beta)P)\\
2R&\leq & \log(1+g_1P) + \log(1+ g_2P) \end{array}\right. .$$
\end{itemize}

Let us focus on the event $\omega_3$: it is easy to prove that
\begin{eqnarray*}
  \omega_3&=&\left\{g_1\leq \frac{2^R-1}{(1+2^R(\alpha -1) )^+P}, \right. \\ && \left. g_1\leq \frac{2^R-1}{(1-2^R\alpha)^+P}, g_1\leq \frac{2^{2R}-1}{P}\right\}. 
\end{eqnarray*}

Let us focus on the event $\omega_3'$: it is easy to prove that
\begin{eqnarray*}
  \omega_3'&=&\left\{ g_2 \geq \left(\max \left( \frac{1}{(1-\beta)P}\left(\frac{2^R}{1+ g_1(1-\alpha)P} -1\right) , \right.\right.\right.\\ && \left.\left.\left. \frac{1}{\beta P}\left(\frac{2^R}{1+g_1\alpha P}-1\right),\frac{1}{P}\left(\frac{2^{2R}}{1+g_1P}-1\right)\right)\right)^+\right\}.
               \end{eqnarray*}

Consequently, if we denote by
\begin{eqnarray*}
  h_3(g_1)&=& \left(\max \left( \frac{1}{(1-\beta)P}\left(\frac{2^R}{1+ g_1(1-\alpha)P} -1\right) \right.\right. ,\\ && \left.\left. \frac{1}{\beta P}\left(\frac{2^R}{1+g_1\alpha P}-1\right),\frac{1}{P}\left(\frac{2^{2R}}{1+g_1P}-1\right)\right)\right)^+
  \end{eqnarray*}
and by
$$G_{\max}=\min \left( \frac{2^R-1}{(1+2^R(\alpha -1) )^+P}, \frac{2^R-1}{(1-2^R\alpha)^+P}, \frac{2^{2R}-1}{P} \right)  $$
we have
$$\mathcal{P}_3(\alpha,\beta)=\frac{1}{\sigma^2} \int_{0}^{G_{\max}} e^{-\frac{h_3(g)}{\sigma^2}} e^{-\frac{g}{\sigma^2}}   dg. $$

\subsection{Closed-form expressions for $\mathcal{P}_4$ and $\mathcal{P}_4'$}
The event $\Omega_4$ corresponds to the case where $(m_1,m_2)$ not decoded at timeslot 1, and $m_1$ (but not $m_2$) decoded at timeslot $2$. Consequently 
$$\mathcal{P}_4=\textrm{Pr}(\omega_3,\omega_4)$$
where $\omega_4$ is the event of success for message $m_1$ and of failure for message $m_2$ when the available observations are  
$$ \left[\begin{array}{c} \mathbf{y}(1) \\ \mathbf{y}(2) \end{array} \right] =  \left[\begin{array}{cc} \sqrt{\alpha P} g_1 & \sqrt{(1-\alpha)P}g_1 \\ \sqrt{\beta P} g_2 & \sqrt{(1-\beta)P}g_2 \end{array} \right] \left[\begin{array}{c} \mathbf{p}(1) \\ \mathbf{p}(2) \end{array} \right]    + \textrm{noise}. $$
The rate region of this channel (corresponding to success for message 1 and to failure of message 2) leads to the two equations 
$$\left\{\begin{array}{lcl} 
R&\leq&  \log(1+\frac{g_1\alpha P}{1+g_1(1-\alpha)P}) + \log(1+\frac{g_2 \beta P}{1+g_2(1-\beta)P}) \\
R&\geq& \log(1+g_1(1-\alpha)P)+ \log(1+g_2(1-\beta)P).\end{array}\right. . $$
So the event $\omega_4$ can be written as follows
$$\left\{ \begin{array}{clc} g_2&\geq &  \frac{ \left(\frac{2^R}{1+\frac{g_1\alpha P}{1+g_1(1-\alpha )P}}-1 \right)^+ }{P\left(  \beta -(1-\beta ) (\frac{2^R}{1+\frac{g_1\alpha P}{1+g_1(1-\alpha )P}}-1)  \right)^+ }\\ g_2 & \leq & \frac{1}{(1-\beta) P }\left(\frac{2^R}{1+g_1(1-\alpha )P}-1 \right)^+  \end{array}\right. .$$
So we denote
$$h_4(g_1;\alpha,\beta)= \frac{ \left(\frac{2^R}{1+\frac{g_1\alpha P}{1+g_1(1-\alpha )P}}-1 \right)^+ }{P\left(  \beta -(1-\beta ) (\frac{2^R}{1+\frac{g_1\alpha P}{1+g_1(1-\alpha )P}}-1)  \right)^+ }$$
and
$$\overline{h}_4(g_1;\alpha,\beta)=\frac{1}{(1-\beta) P }\left(\frac{2^R}{1+g_1(1-\alpha )P}-1 \right)^+.$$

If $h_4> \overline{h}_4$, then the event on $g_2$ is never true and does not take part on $\mathcal{P}_4$.
If not, it takes part.
Therefore, we have 
$$\mathcal{P}_4(\alpha,\beta)=\frac{1}{\sigma^2} \int_{0}^{G_{\max}}  \left( e^{-\frac{h_4(g;\alpha,\beta)}{\sigma^2}} - e^{-\frac{\overline{h}_4(g;\alpha,\beta)}{\sigma^2}}  \right)^+  e^{-\frac{g}{\sigma^2}}  dg.$$

As for $\mathcal{P}_4'$, we can replaced $\alpha$ with $1-\alpha$ and $\beta$ with $1-\beta$.
Indeed the event $\omega_3$ is symmetric with respect to the change of variable $\alpha\to 1-\alpha$ since both messages with power $\alpha P$ and $(1-\alpha)P$ need to fail. And for the second timeslot, we permute the role between message $1$ and message $2$. Therefore, we have 
$$\mathcal{P}_4'(\alpha,\beta)=\mathcal{P}_4(\alpha,1-\beta).$$

\subsection{Closed-form expressions for $\tilde{\mathcal{P}}_3$, $\tilde{\mathcal{P}}_4$, and $\tilde{\mathcal{P}}_5$}
As $\tilde{\Omega}_3$ corresponds to the event of $\Omega_3$ only occurring in timeslot 2. We have $\tilde{\Omega}_3=\omega_3'$ which implies that $ \tilde{\mathcal{P}}_3 = \textrm{Pr}(\omega_3')$. We straightforwardly deduce that 
\begin{eqnarray*}
  \tilde{\mathcal{P}}_3(\alpha)
  &=&\frac{1}{\sigma^2} \int_{0}^{\infty} e^{-\frac{h_3(g)}{\sigma^2}} e^{-\frac{g}{\sigma^2}} dg.
  \end{eqnarray*}
By similar argument, we have $\tilde{\mathcal{P}}_4=\textrm{Pr}(\omega_4)$, which implies that 
$$\tilde{\mathcal{P}}_4(\alpha)=\frac{1}{\sigma^2} \int_{0}^{\infty}  \left( e^{-\frac{h_4(g;\alpha,\alpha)}{\sigma^2}} - e^{-\frac{\overline{h}_4(g;\alpha,\alpha)}{\sigma^2}}  \right)^+  e^{-\frac{g}{\sigma^2}}   dg . $$
Moreover it is easy to prove that 
$$\tilde{\mathcal{P}}_4'(\alpha)=  \tilde{\mathcal{P}}_4(1-\alpha).$$


\section{Numerical results}\label{sec:simus}
In this Section, we numerically evaluate the closed-form expressions obtained for the throughput of the different protocols. We define the \ac{SNR} as $P/\sigma^2$. 

\begin{figure}[htb]
\begin{center}
\includegraphics[width=0.8\columnwidth]{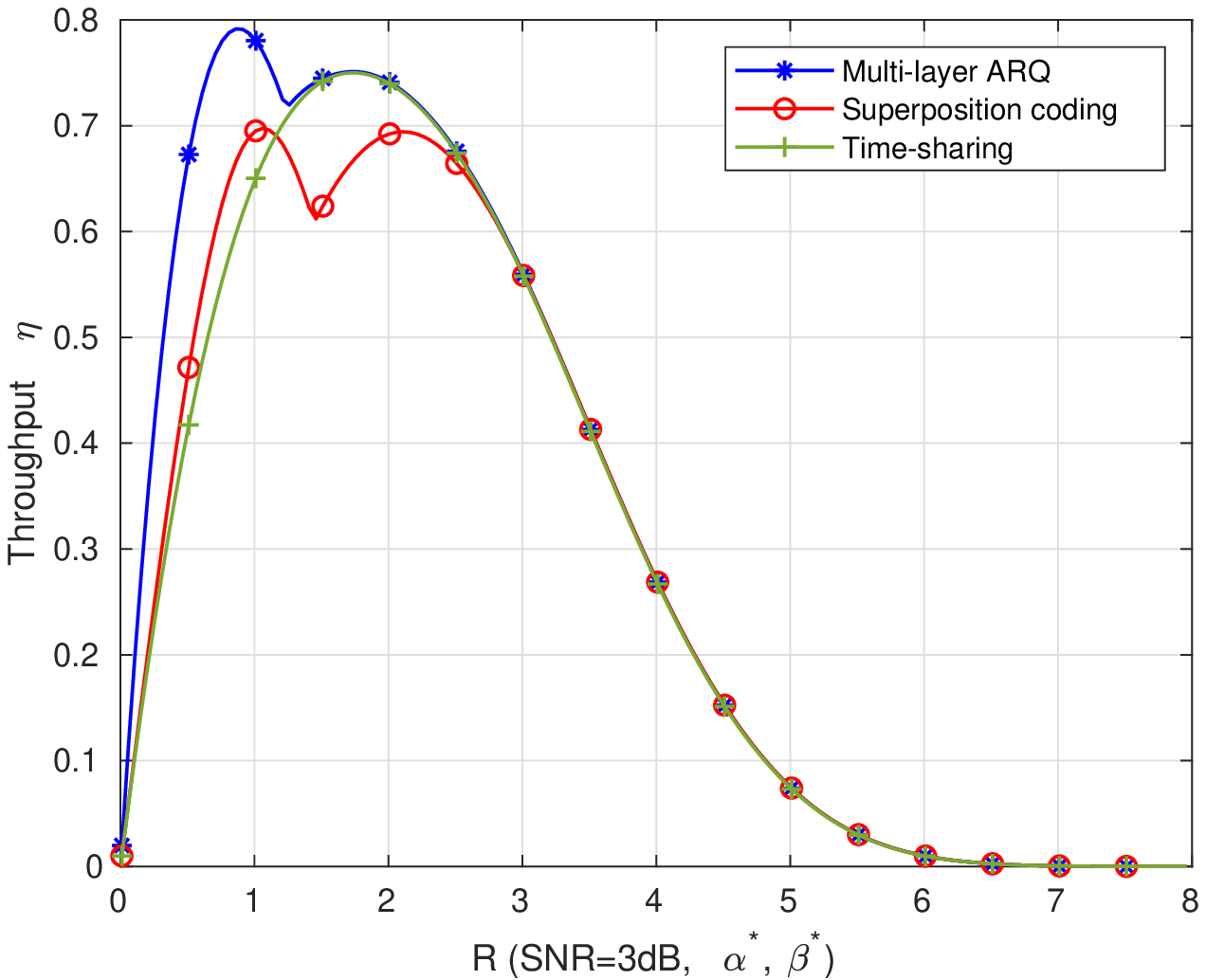}
\caption{Throughput versus $R$ (with SNR$=3$dB, $\alpha^*$, and $\beta^*$)}\label{fig:T_vsRSNRa*}
\end{center}
\end{figure}

In Fig.~\ref{fig:T_vsRSNRa*}, we plot the throughput versus $R$ for \ac{SNR}$=3$dB and the optimized $\alpha$ and $\beta$. The $\alpha$ and $\beta$ are optimized for each $R$ and each protocol. 
Clearly, the performance are better by applying a multi-layer \ac{HARQ} for small $R$. Indeed, when $R$ is small enough the probability to decode in one timeslot is high and sending the packets into two parts (with feedback) is beneficial compared to the superposition coding over both timeslots. In addition, the time-sharing is worse than the well-tuned superposition coding for small $R$. 

\begin{figure}[htb]
\begin{center}
\includegraphics[width=0.8\columnwidth]{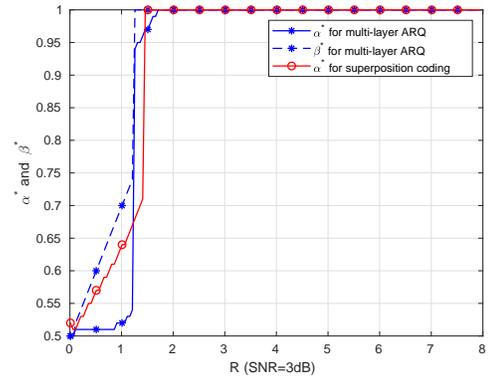}
\caption{$\alpha^*$ and $\beta^*$ versus $R$ (with SNR$=3$dB)}\label{fig:a*_vsRSNR}
\end{center}
\end{figure}

In Fig.~\ref{fig:a*_vsRSNR}, we plot the optimized $\alpha^\star$ and $\beta^\star$ (maximizing the throughput) versus $R$. 
We see that when $R$ increases, the optimal value of $\alpha$ and $\beta$ are $1$ which correspond to the standard parallel \ac{HARQ}. Actually when $R$ is large, the probability to be correctly received in one timeslot is very small and the interference induced by the multiplexing comes a serious drawback and it is better to have a conservative policy with only one message at each timeslot.  

\begin{figure}[htb]
\begin{center}
\includegraphics[width=0.8\columnwidth]{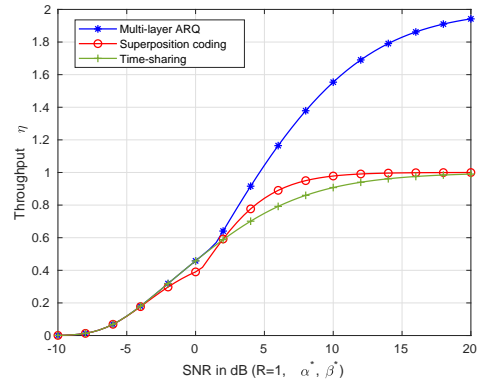}
\caption{Throughput versus SNR (with $R=1$, $\alpha^*$, and $\beta^*$)}\label{fig:T_vsSNRRa*}
\end{center}
\end{figure}
\begin{figure}[htb]
\begin{center}
\includegraphics[width=0.8\columnwidth]{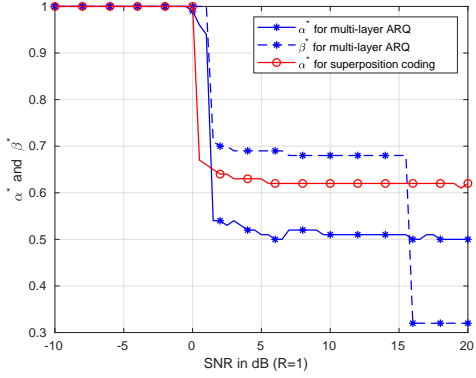}
\caption{$\alpha^*$ and $\beta^*$ versus SNR (with $R=1$)}\label{fig:a*_vsSNRR}
\end{center}
\end{figure}

In Fig.~\ref{fig:T_vsSNRRa*}, we plot the throughput versus \ac{SNR} for $R=1$ and the optimized $\alpha$ and $\beta$. The $\alpha$ and $\beta$ are optimized for each \ac{SNR} and each protocol and are given in Fig. \ref{fig:a*_vsSNRR}. 
The throughput for multi-layer \ac{HARQ} is slightly higher for mid-\ac{SNR} and dramatically higher for large \ac{SNR}. When the link is very poor, it is better to send the whole packet within both timeslots.  The superposition really operates when the \ac{SNR} is high enough, i.e., when the link is reliable enough. Then, both packets of the multi-layer \ac{HARQ} protocol require only $1$ timeslot to be successfully decoded which leads to an asymptotic value of $2$. In contrast, both time-sharing and superposition coding need $2$ timeslots to transmit $2$ packets of rate $R$. The standard protocols are penalized  but they may counteract by using another value of the rate, for instance, by forcing their rates to be $2R$ where $R$ is the rate of each message of the multi-layer protocol.

\begin{figure}[htb]
\begin{center}
\includegraphics[width=0.8\columnwidth]{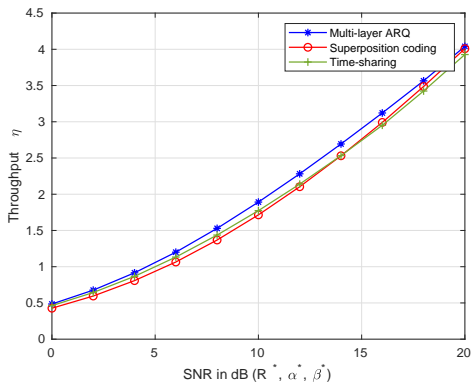}
\caption{Throughput versus SNR (with $R^*$, $\alpha^*$, and $\beta^*$)}\label{fig:T_vsSNRR*a*}
\end{center}
\end{figure}

\begin{figure}[htb]
\begin{center}
\includegraphics[width=0.8\columnwidth]{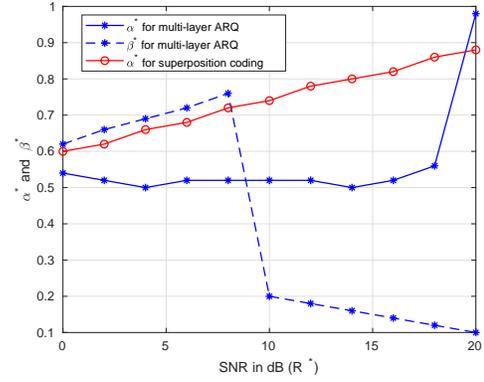}
\caption{$\alpha^*$ and $\beta^*$ versus SNR (with $R^*$)}\label{fig:a*_vsSNRR*}
\end{center}
\end{figure}

\begin{figure}[htb]
\begin{center}
\includegraphics[width=0.8\columnwidth]{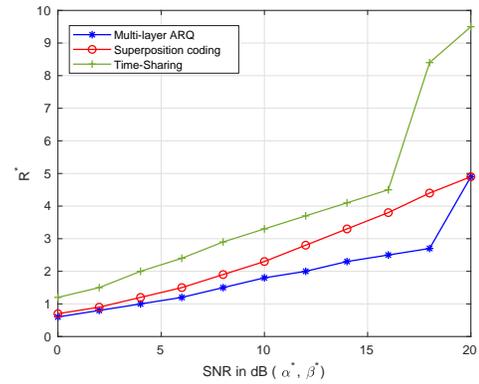}
\caption{$R^*$ versus SNR (with $\alpha^*$ and $\beta^*$)}\label{fig:R*_vsSNRa*}
\end{center}
\end{figure}

In Fig.~\ref{fig:T_vsSNRR*a*}, we plot the throughput versus \ac{SNR} for the optimized $R$, $\alpha$, and $\beta$. For each \ac{SNR} and each protocol, $R$, $\alpha$, and $\beta$ are optimized, and are given in Figs.~\ref{fig:a*_vsSNRR*} and \ref{fig:R*_vsSNRa*} respectively. 
Tuning well parameters $R$, $\alpha$, and $\beta$ enable standard protocols to be much closer to the multi-layer \ac{HARQ}. Nevertheless, at mid-\ac{SNR}, the gains in throughput are about $11\%$ for multi-layer \ac{HARQ} protocol compared to the superposition coding and about $7\%$ compared to the time-sharing. 


\section{Conclusion}\label{sec:ccl}
We have analyzed the performance of a multi-layer \ac{HARQ} compared to the standard \ac{HARQ} and the superposition coding. For a given rate, the gain in throughput may be substantial. For an optimized rate, the gain is only noticeable at mid-\ac{SNR}.

\end{document}